\documentclass[journal=jacsat,manuscript=article]{achemso}

\usepackage[version=3]{mhchem} 
\usepackage{graphicx} 
\usepackage{placeins}
\usepackage[normalem]{ulem}

\author{Daphn\'e Lubert-Perquel}
\affiliation{Chemistry and Nanoscience Center, National Renewable Energy Laboratory, 15013 Denver West Pkwy, Golden, CO 80401}
\author{Byeong Wook Cho}
\affiliation{Center for Integrated Nanostructure Physics, Department of Energy Science \& Department of Physics, Sungkyunkwan University, Suwon 16419, Korea}
\author{Alan J. Philips}
\affiliation{Chemistry and Nanoscience Center, National Renewable Energy Laboratory, 15013 Denver West Pkwy, Golden, CO 80401}
\author{Young Hee Lee}
\affiliation{Center for Integrated Nanostructure Physics, Department of Energy Science \& Department of Physics, Sungkyunkwan University, Suwon 16419, Korea}
\author{Jeffrey L. Blackburn}
\author{Justin C. Johnson}
\email{justin.johnson@nrel.gov}
\affiliation{Chemistry and Nanoscience Center, National Renewable Energy Laboratory, 15013 Denver West Pkwy, Golden, CO 80401}

\title[An \textsf{achemso} demo]
  {Modulating spin-valley relaxation in WSe$_2$ with variable thickness VOPc layers}

\keywords{American Chemical Society, \LaTeX}

\begin{document}

\begin{abstract}

Combining the synthetic tunability of molecular compounds with the optical selection rules of transition metal dichalcogenides (TMDC) that derive from spin-valley coupling could provide interesting opportunities for the readout of quantum information. However, little is known about the electronic and spin interactions at such interfaces and the influence on spin-valley relaxation. In this work, vanadyl phthalocyanine (VOPc) molecular layers are thermally evaporated on WSe$_2$ to explore the effect of molecular layer thickness on excited-state spin-valley polarization. The thinnest molecular layer supports an interfacial state which destroys the spin-valley polarization almost instantaneously, whereas a thicker molecular layer results in longer-lived spin-valley polarization than the WSe$_2$ monolayer alone. The mechanism appears to involve a tightly-bound species at the molecule/TMDC interface that strengthens exchange interactions and is largely avoided in thicker VOPc layers that isolate electrons from WSe$_2$ holes.

\end{abstract}

\section{Introduction} 

Read-out of stored quantum information is a key requirement of all spin qubits, and with the advent of 2D materials such as graphene or transition metal dichalcogenides (TMDCs), the ability to exploit the local energy extrema in momentum space known as valleys opens possibilities to emergent “valleytronic” concepts.\cite{liu_2d_2019, Pal2023} In TMDCs, inequivalent valleys at the ±K points of the Brillouin zone act as pseudo-spins of opposite polarization. Moreover, direct band gap TMDC monolayers have demonstrated circular polarized photoluminescence (PL) from valley-selective excitation and emission.\cite{Zeng2012, Mak2012}  As chiral emission and spin-photon interfaces are relevant to spin-selective optoelectronics, combining a molecular spin qubit with a TMDC substrate could lead to interesting opportunities in quantum information science (QIS). Molecular systems are the most recent candidates for QIS, with several recent perspectives and reviews discussing the valuable properties unique to such systems.\cite{Troiani2019, Atzori2019, Wasielewski2020, reid_molecular_2024} However, at this stage little is known of the electronic and spin interactions at the interface between molecules and 2D materials and the mechanisms of spin-valley relaxation.  

Recent work on WSe$_2$ / C$_{60}$ bilayers has shown that interfacial charge separation after photoexcitation of the TMDC can prolong the spin-valley polarization time.\cite{Zhao2021} The proposed mechanism involves charge hopping away from the interface after the initial charge-transfer event, leading to reduced exchange interactions that otherwise serve to hasten spin-valley polarization decay.   The specific properties of the bilayer (particularly the molecular ordering and interaction with the WSe$_2$ orbitals) and the degree to which the polarization time can be prolonged remain unknown. More controlled molecular systems with tunable electronic structure and well-defined orientations in thin-film assemblies may be advantageous toward answering these fundamental questions. The molecular system considered in our study is vanadyl phthalocyanine (VOPc). Depositing this on WSe$_2$ is expected to form a type II heterojunction, similar to that reported for the WSe$_2$/C$_{60}$. Prior work with VOPc deposited on MoS$_2$ has shown conflicting results, with some suggestion of mid-gap trap state formation,\cite{Kong2022} and in other cases indirect emission across the interface with charge-transfer (CT) character.\cite{Schwinn2022}  

Here we present a study of VOPc, a well-known spin system with potential utility in quantum sensing schemes, thermally evaporated on WSe$_2$ monolayers grown by chemical vapor deposition (CVD). The modification of spin-valley polarization lifetimes as a function of molecular layer thickness shows that this tunable bilayer system can balance strong interfacial interactions (thin VOPc layers) with morphology control (thicker layers), which is important for enhancing communication between the molecular spin and the optically active WSe$_2$ pseudospin. Whereas the bilayer with 10 nm thickness of VOPc appears to extend spin-valley polarization, as observed previously for the WSe$_2$-C$_{60}$ system, a thinner 5 nm VOPc layer destroys the spin-valley polarization present in the individual WSe$_2$ layer on an ultrafast timescale. Several hypotheses are put forward to explain the distinction in behavior as a function of molecular thickness, including the formation of a hybrid exciton, and the interplay between strongly and weakly bound charge-transfer states. These results provide critical information to design new architectures for quantum devices that may leverage optical signatures of versatile spin systems.

\section{Results and Discussion}

\begin{figure}[h]
  \centering
    \includegraphics[scale=0.8]{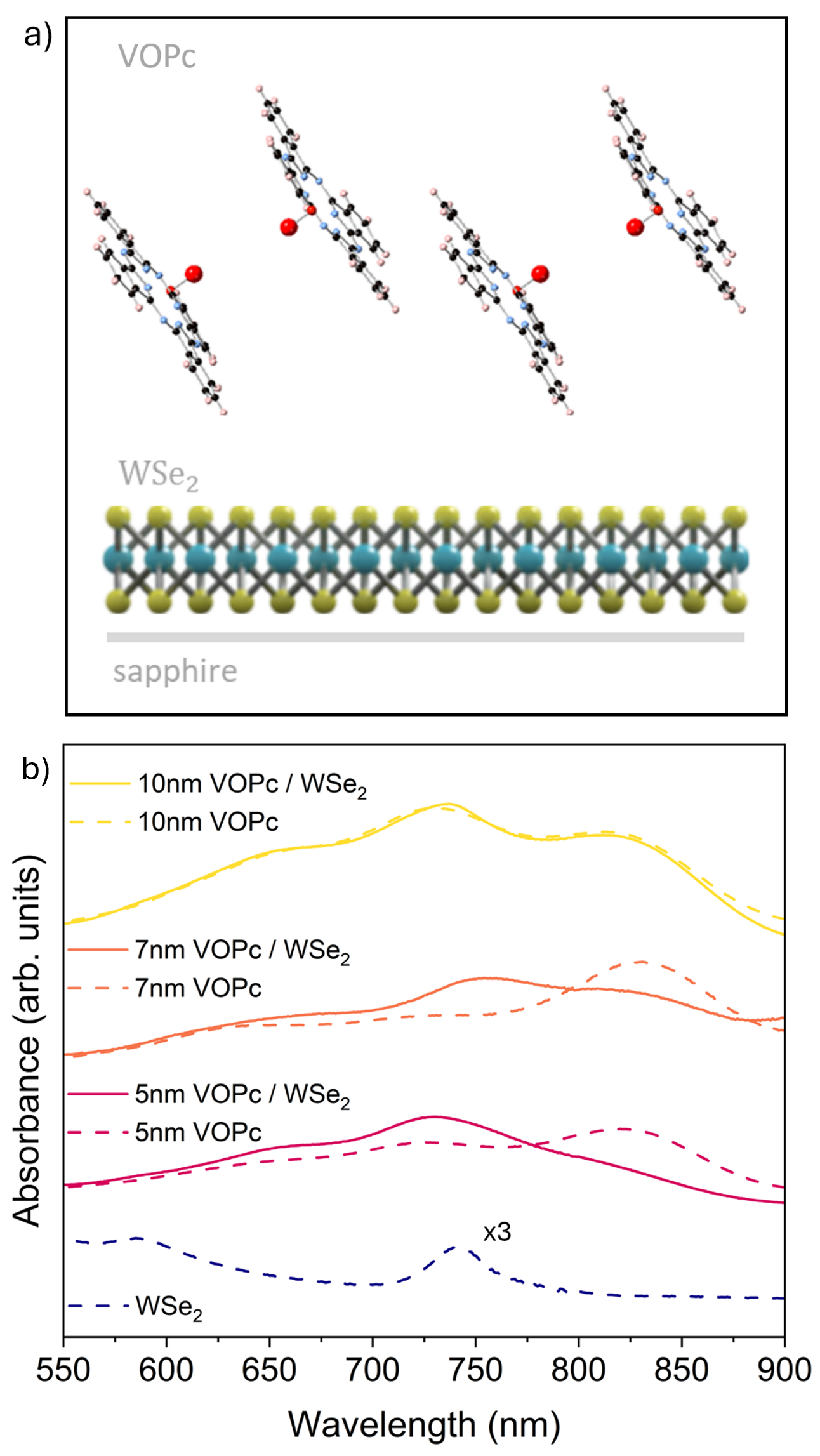}
  \caption{a) Schematic of the VOPc molecules on the WSe$_2$, as determined from XRD of the 10 nm VOPc layer. b) Absorbance spectra of VOPc and WSe$_2$ for the individual and bi-layers. The pure WSe$_2$ trace is weak as originating from a monolayer and is therefore magnified by a factor 3 to compare on the same scale as the other layers.}
  \label{Fig1}
\end{figure}

VOPc was thermally evaporated on monolayer flakes of WSe$_2$ grown by CVD and transferred onto custom-sized sapphire substrates (detailed methods provided in the supplementary information). The molecules are found to deposit at an angle of $\sim$70$^{\circ}$ between the molecular plane and substrate, schematically represented in Figure \ref{Fig1}a. This angle was determined from the peak observed by x-ray diffraction (XRD) in the 10~nm VOPc on WSe$_2$ sample (Figure S3).\cite{Liu2019} Thinner samples did not have a measurable XRD signal and therefore the orientation of the molecules can not be verified. Polarized Raman spectroscopy can be used to compare the molecular orientation of the VOPc / WSe$_2$ bilayers.\cite{Basova1998, xiong2024} In this experiment, Raman spectra are taken with the excitation polarization parallel and orthogonal to the detection polarization, Figure S4. The larger ratio of the parallel/orthogonal peaks associated with pyrrole stretching in the 5nm VOPc bilayer confirms increased tilting of the molecules toward the WSe$_2$ surface.

The absorbance spectra of the pure WSe$_2$ monolayer, 5~nm, 7~nm and 10~nm VOPc, as well as the three thicknesses of bilayers are shown in Figure \ref{Fig1}b. The pure monolayer WSe$_2$ has the characteristic A and B excitons that acquire oscillator strength from the split valence bands at 730~nm and 590~nm respectively. Similarly, the VOPc spectra show the Q-band with three sub-features indicative of the phase II polymorph. Information can be gleaned on the packing and orientation of molecules from the ratio of Q-band features.\cite{Griffiths1976, Wang2014} The relative intensity of the 0-0 and 0-1 vibronic peaks is an indication of the local ordering in these polycrystalline films. In the 5~nm and 7~nm VOPc on sapphire, the 830~nm peak is strongest, confirming the dominance of polymorph II.  Substrate-induced alterations are largest for the 5 nm thick VOPc film on WSe$_2$ with the quenching of the low energy peak. It is clear that the peak positions are unchanged between the samples, though the relative intensity of the 0-0 and 0-1 peaks vary, and as such is attributed to a more amorphous structure, which would also explain the lack of XRD signal. This evidence of reduced ordering likely originates from interfacial interactions between VOPc and WSe$_2$ that encourage a more cofacial structure as seen in  polymorph I.\cite{delCano2005, Wang2014} However, due to limitations from the sample size and transparent substrate, structural information could not be obtained for the thin films.  Increasing the VOPc thickness to 10~nm shows a stronger 710~nm peak indicating local disorder from poorer crystallinity or possibly from the presence of both polymorphs. It should be noted that the absorbance of the 10 nm VOPc film  is largely unchanged, whether deposited onto sapphire or onto WSe$_2$, confirming that for thicker layers the VOPc ordering is not impacted substantially by the underlying substrate. In this case the growth is dominated by intermolecular interactions. To summarise, the thin films of VOPc on WSe$_2$ appear to have a more amorphous interface with reordering occurring in thicker films as the intermolecular interactions dominate over the interfacial ones.

\begin{figure*}[h]
 \centering
 \includegraphics[height=9cm]{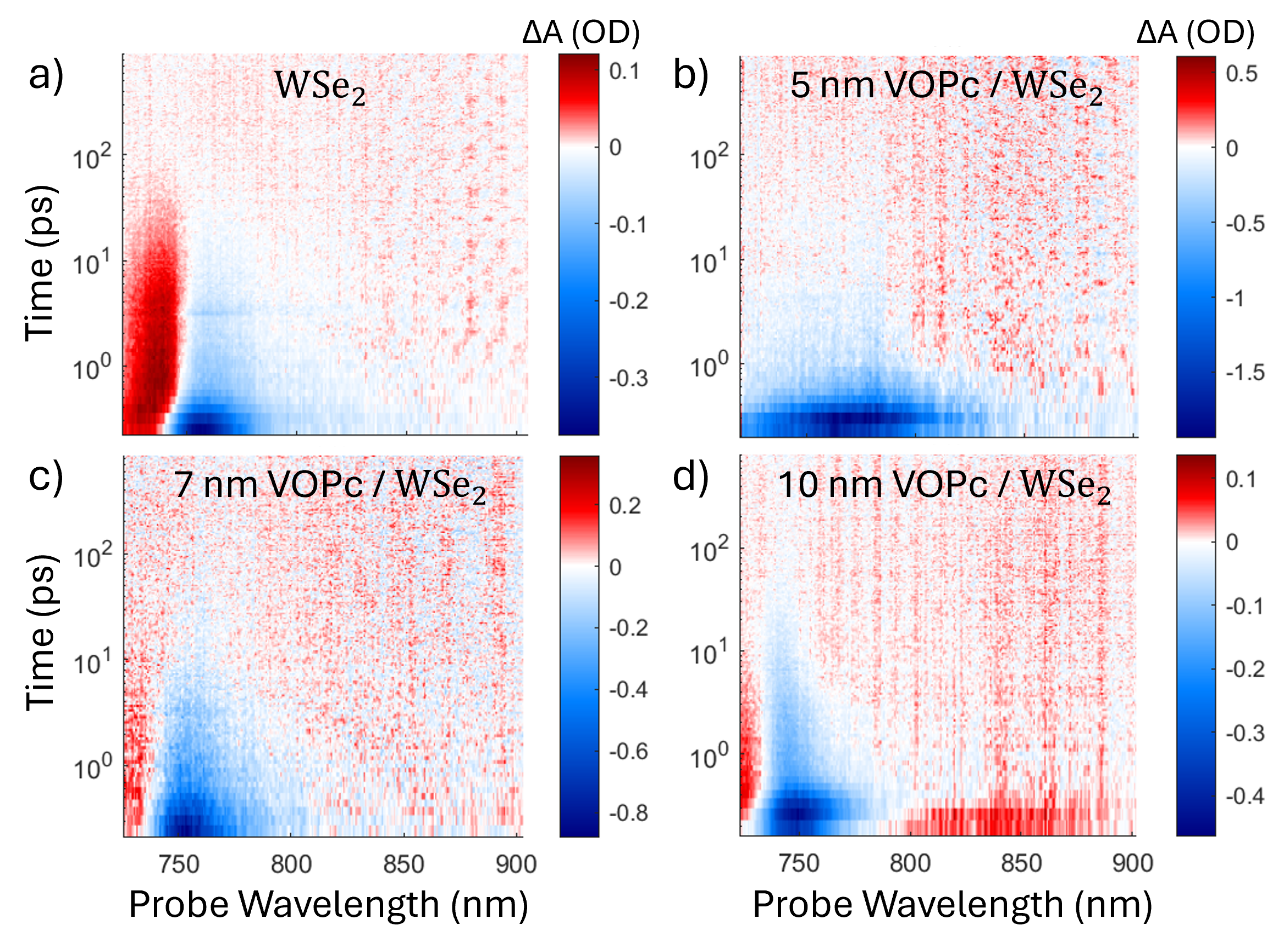}
 \caption{2D plots of the time-dependent TRCD (SP-OP) spectra after 710 nm excitation for the a) WSe$_2$, b) 5~nm VOPc on WSe$_2$, c) 7~nm VOPc on WSe$_2$, d) 10~nm VOPc on WSe$_2$. The 5~nm sample, unlike other samples, has a broader red-shifted signal. The A-exciton feature in the 7~nm and 10~nm matches that observed in the pure WSe$_2$ with an additional photoinduced absorption feature in the 10~nm bilayer corresponding to the VOPc contribution. }
 \label{fgr:TRCD}
\end{figure*}

In recent years, molecular systems have been put forward as an alternative for quantum applications due to the tunability of their physical properties.\cite{Bayliss2020, Wasielewski2020, Atzori2019} Simple parameters such as temperature and magnetic field are common tools to manipulate the ground-state electron spin population required for these processes. VOPc is considered here as the molecular candidate, but the spin population has no convenient readout mechanism. One of the primary reasons that TMDCs are interesting is for their spin-valley coupling revealed through circularly-polarized emission,  so combining these two materials could provide a system with a measurable response to the tunable spin population. However, first the effect of the molecular layer on the spin-valley coupling of the TMDC needs to be understood. This is incisively investigated using time-resolved circular dichroism (TRCD) spectroscopy and supported by transient absorption (TA) spectroscopy. Features of the transient spectra can be interpreted by comparing the spectral shapes of kinetic traces as detailed in literature.\cite{Schiettecatte2023} TRCD is a pump-probe technique with polarizing optics in both the pump and probe beam paths. A brief explanation is given in the supplementary information, but further detail can be found in a previously published review.\cite{lubert2023} The spin polarization is determined by comparing the transient spectra where the pump and probe have the same circular polarization (SP) and opposite circular polarization (OP), enabling detection of population residing either in the same or opposite spin valley (Figure S1). The kinetic fit of the difference is used to estimate the spin-valley relaxation time (Figure S2). We note that our kinetic results are shown as SP-OP, although this does not truly isolate the degree of spin-valley polarization as it is convoluted with population relaxation.\cite{Wangsciadv} However, deconvoluting these given limited signal-to-noise ratios was not possible but does not affect the trends observed.

The pure WSe$_2$ was measured first as a benchmark, Figure 2a. The lineshape reflects the exciton absorption profile and as such the SP spectra contain a simple bleach due to phase-space filling. The OP spectra are not expected to contain a phase-space filling component and indeed the lineshape are dispersive instead, Figure S2. This reflects a shift in the exciton absorption profile in the unpumped valley that is interrogated by the oppositely polarized probe.\cite{Zhao2021} Moreover, in agreement with literature, the kinetic fit of the SP$-$OP, Table 1, is a biexponential with decay parameters $\tau_1$~=~200~fs corresponding to the intervalley scattering through the electron-hole exchange interaction and a slower decay, $\tau_2$~=~5~ps, which will be discussed in more detail below. 

\begin{table}[]
\caption{Global fit parameters for the SP-OP traces at 710 nm excitation for the 0, 5, 7 and 10 nm VOPc on WSe$_2$.}
\begin{tabular}{|l|l|l|l|}
\hline
      & $\tau_1$            & $\tau_2$           & $\tau_3$            \\ \hline
0 nm  & 0.2 $\pm$ 0.004 ps  & 5.0 $\pm$ 0.15 ps    &                     \\ \hline
5 nm  & 0.37 $\pm$ 0.006 ps &                    &                     \\ \hline
7 nm  & 0.30 $\pm$ 0.02 ps  & 4.94 $\pm$ 0.28 ps & \textgreater 100 ps   \\ \hline
10 nm & 0.31 $\pm$ 0.007 ps & 6.85 $\pm$ 0.22 ps & \textgreater 100 ps \\ \hline
\end{tabular}
\label{GF params}
\end{table}

The WSe$_2$ / VOPc bilayers similarly show the bleach in SP and dispersive shape in OP spectra for the A-exciton feature, resulting in the similar pattern of features evident in Figure 2b-d. However, the breadth and persistence of such features, qualitatively evident in the 2D maps, can be further visualized through sliced spectra from these datasets in Figure S5 and spectral components from global fits in Figure 3. From the individual spectra it can be seen that the VOPc bleach possesses no circular dichroism at any delay time. It therefore acts as an internal standard that references the relative SP and OP differences at the A-exciton feature, shown in Figure 2. Figure 3a-d contain the global fit spectra associated with the kinetic components. Notably, considerable variations in the kinetic parameters of the spin-valley relaxation are obtained for the 5~nm vs. 7~nm \& 10~nm VOPc bilayers, Table \ref{GF params}. In the thinnest 5 nm film (Figure 3b), the ultra-fast component associated with intervalley scattering is present, but the longer-lived component is not. The TRCD spectral features are also considerably broader and somewhat red-shifted in the 5 nm bilayer compared to all other samples, as is also clearly evident in Figure 2b. It is proposed that the WSe$_2$ and VOPc layers interact in this sample such that they engender a tightly-bound interlayer state. True hybridization of the electronic states of the layers would perturb the band alignment, resulting in broken symmetry that nullifies spin-valley polarization regardless of the initial valley polarization induced from the circularly polarized pump.\cite{Amit2023} Broadening and quenching of slower spin-valley polarization decay times are evidence of the strong perturbation of this interfacial state. We provide further characterization of this state below.

\begin{figure*}
 \centering
 \includegraphics[height=9cm]{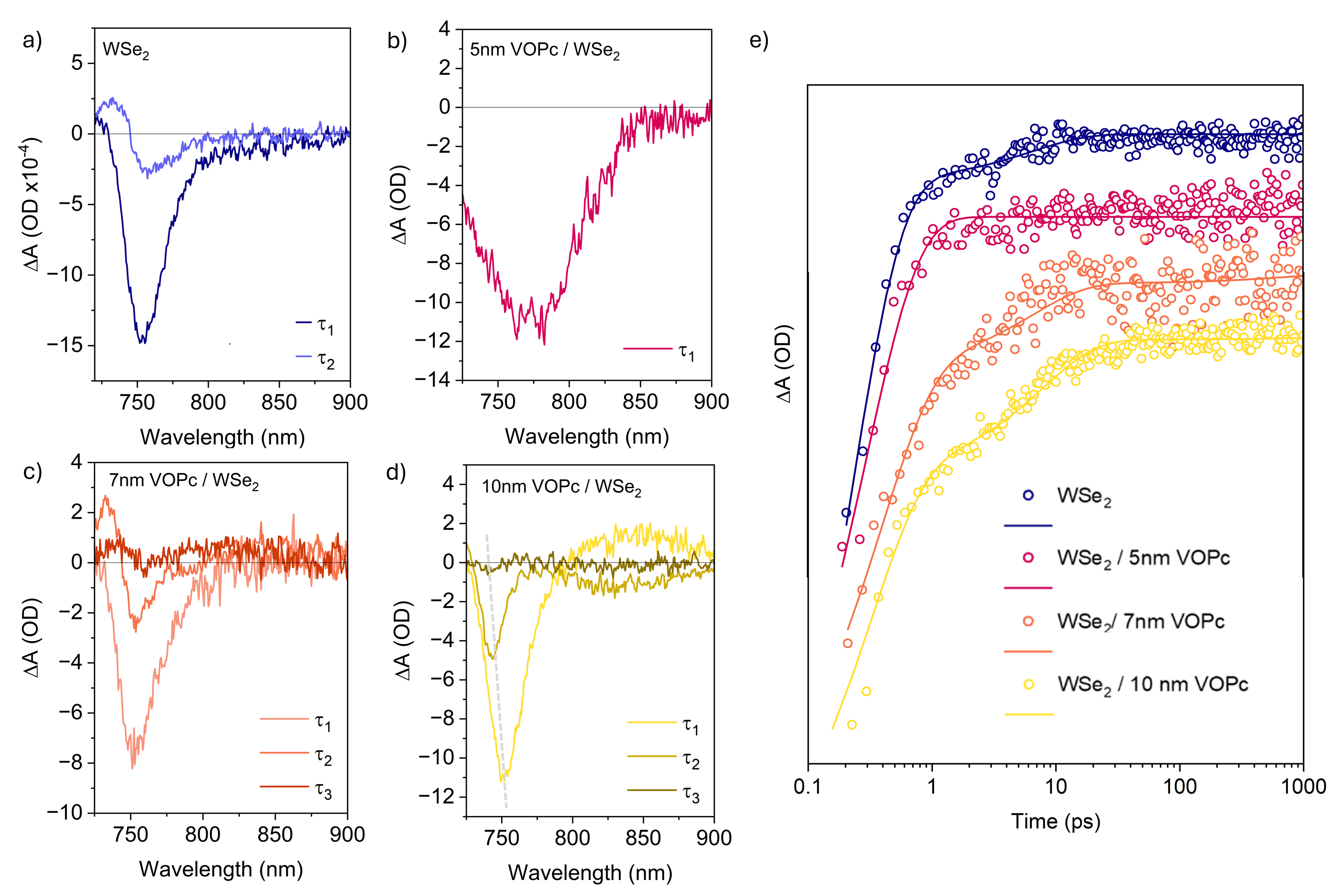}
 \caption{a)-d) Decay-associated spectra obtained from a global fit for WSe$_2$, 5 nm VOPc/WSe$_2$, 7 nm VOPc/WSe$_2$ and 10 nm VOPc/WSe$_2$ respectively. $\Delta A$ is defined as the change in absorbance (synonymous with the change in optical density). The grey dotted line in panel (d) is a guide to the eye to see the decrease in amplitude and blue-shift of the spectral shape between parameters.  e) Comparison of the kinetic traces of the WSe$_2$ and VOPc bilayers at the peak intensity (750-755 nm) for each sample (open circles) with the global fit traces (solid lines). The pure WSe$_2$ is scaled for a clearer comparison to the other samples.}
 \label{fgr:TRCDb}
\end{figure*}

On the other hand, the 7~nm and 10~nm VOPc bilayers (Figure 3c, d) show both kinetic parameters observed in the pure WSe$_2$ with little spectral perturbation.  However, a weak but non-negligible long-lived component ($>$ 100 ps) with a dispersive line shape appears for both (yellow traces, Figure 3c,d), indicating a red shift of the A-exciton is associated with the slowly decaying spin-valley polarization.\cite{Schiettecatte2023}  The amplitude of this component also appears to be proportional to the thickness of the VOPc in the sample. Noise at later delay times prohibits determination of a lifetime with reasonable uncertainty bounds, and we assign its value simply as greater than 100 ps. We propose that photoinduced charge transfer occurs between the layers, creating a stark-shifted TRCD signal that persists because the hole that remains in the WSe$_2$ after charge transfer is partially protected from spin-valley relaxation through reduced exchange interactions\cite{crooker}.  This long-lived component has been seen previously in bilayers of WSe$_2$ and C$_{60}$.\cite{Zhao2021}

The decay associated spectra from the global fits, Figure 3a-d, reveal the primary decay related to the A-exciton bleach, and the dispersive intermediate decay component. The spectral shape of this second component is postulated to result from an effective trion absorption that results from probe interrogation of an excitation in the K valley interacting with a hole created in the K' valley with the oppositely polarized pump.\cite{Wangsciadv}  Though small in relative amplitude, the slowest-decaying components appear to have the opposite sense of spectral shifting, suggesting a mechanism related to Stark shifting of the A-exciton due to long-range charge separation.\cite{stark} Figure 3e demonstrates the change in spin-valley polarization decay through kinetic slices at the A-exciton bleach and the associated bi- or triexponential fit curves. Additional graphs are provided in Figure S6 to clarify the comparisons between samples.

\begin{figure}[h]
    \includegraphics[scale=0.75]{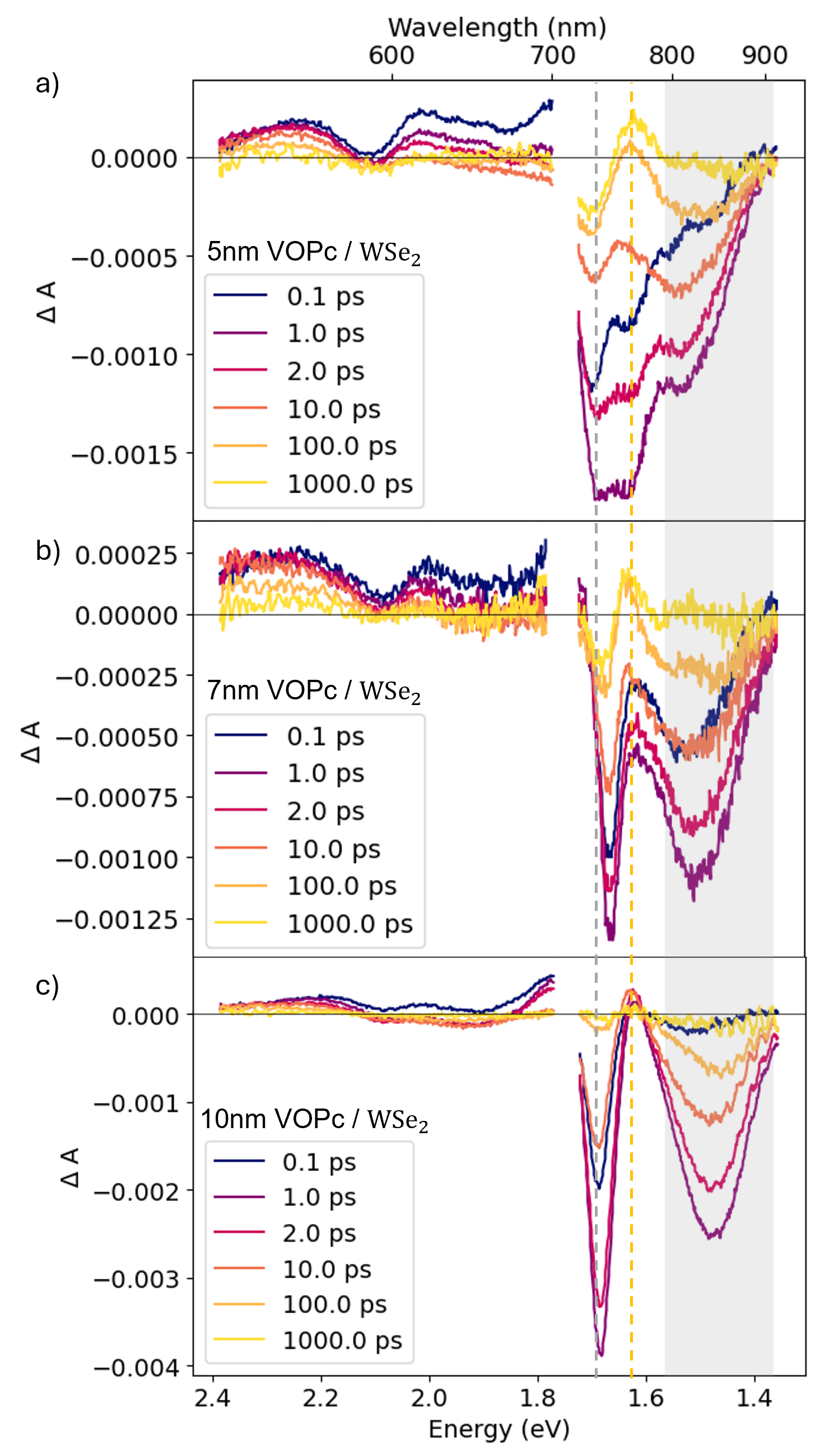}
  \caption{Transient absorption spectra with excitation wavelength 710~nm  at a fluence of 20 $\mu$J/cm$^2$ for a) 5~nm VOPc on WSe$_2$, b) 7~nm VOPc on WSe$_2$ and c) 10~nm VOPc on WSe$_2$ .}
  \label{Bilayer_TA}
\end{figure}

From TA spectroscopy in the 5~nm bilayer (Figure \ref{Bilayer_TA}a), a 1.65 eV shoulder (dashed orange line) to the A-exciton peak (dashed gray line) provides some evidence of an interlayer state. There is also evidence of either charge or energy transfer to VOPc shown by the growth of the VOPc ground state bleach at $\sim$1.5~eV (shaded gray areas), concomitant with loss of WSe$_2$ A-exciton bleach. Selectively exciting the VOPc at 820~nm (Figure S7) results in a long-lived feature at 1.7~eV corresponding to the WSe$_2$ A-exciton. This provides specific evidence of photoinduced charge transfer from VOPc to WSe$_2$. As thermodynamically uphill energy transfer is disallowed under these excitation conditions, charge transfer is the only route to producing the WSe$_2$ ground-state bleach. The charge transfer commences on a sub-ps timescale and leads to a state that persists beyond 1 ns.  The strongly dispersive lineshape of this long-lived feature, absent in the 10~nm bilayer, further supports a long-lived, tightly-coupled, interfacial charge-transfer state.\cite{wang2023}

On the other hand, the 10 nm VOPc bilayer shows parallel decay kinetics when both layers are simultaneously excited (Figure 4c), with some charge transfer occurring relatively slowly due to evidently weak electronic coupling at the interface. Electrons that do cross the interface into VOPc can likely diffuse away from WSe$_2$, leading to the long-lived component in TRCD that corresponds to the slow spin-valley relaxation time of the remaining hole in WSe$_2$.  No charge transfer occurs upon 820 nm excitation, suggesting that the energy level alignment and/or electronic coupling are distinct from that of the 5 nm bilayer.  Interestingly, the 7 nm bilayer exhibits behavior intermediate between the 5 nm and 10 nm samples: a feature near 1.6 eV is present in the transient spectrum, but it is red-shifted from, and less distinct than, that observed for the 5 nm sample. It also becomes more evident at later times.

A 20~nm VOPc on WSe$_2$ was also studied using TA spectroscopy, Figure S8. However, due to the dominant VOPc absorption the A-exciton feature was too weak to measure conclusively by TRCD and was not investigated further. 

\FloatBarrier

From the perspective of the charge transfer and hopping model described in previous literature, one might predict that the spin-valley relaxation time would increase monotonically from the pure WSe$_2$ layer toward thicker VOPc layers, eventually saturating at the thickness associated with the charge diffusion length within VOPc. What we observe is inconsistent with the notion and requires consideration of alternate models to explain the change in photophysical behavior among the 5 nm, 7 nm and 10 nm samples. A first candidate model posits that the different VOPc thicknesses result in varying strengths of electronic coupling at the interface with WSe$_2$, leading to a hybridized state under the strongest coupling conditions.  Prior work has suggested that VOPc can adopt a variety of configurations at a planar interface, depending on deposition conditions and the nature of the underlying substrate.\cite{Wang2014, Cimatti2019} Flat molecular orientations of phthalocyanines are often found in metal and graphene systems\cite{Barlow2000, Niu2014, sessoli_vopc}, and TiOPC was also inferred to lie roughly flat on WSe$_2$.\cite{xiong2024} Moreover, the thickness dependent orientation of CuPc has been reported,\cite{Ruocco2004} leading us to plausibly conjecture that van der Waals forces, though relatively weak, can enable a more parallel alignment of the phthalocyanine ring system with the WSe$_2$ surface for thin VOPc layers. We also qualitatively infer this from polarized Raman experiments, Figure S4.  For thicker VOPc layers, strong intermolecular forces may drive the VOPc layer toward the most stable crystalline configuration, tilting the VOPc molecules to more oblique angles with respect to the WSe$_2$ surface, and negating opportunities for hybrid state formation. This weak coupling regime is likely to result in parallel excited-state kinetics on either side of the WSe$_2$/VOPc interface, similar to what is hypothesized to occur in WSe$_2$/C$_{60}$.\cite{Zhao2021} This situation is reflected schematically in Figure 5a.

Another model relies upon the degree to which charge transfer results in a tightly- vs. weakly bound state.  It was hypothesized in MoS$_2$/VOPc bilayers that tightly-bound species expected to form at the interface are outcompeted by the formation of a highly delocalized ``free charge" state resulting from hot exciton dissociation.\cite{wang2023, friend} When a layer is too thin or highly amorphous, it may be unable to support this free charge state. This type of model is juxtaposed with a related scheme, termed the ``Distributed Range electron transfer model",\cite{Carr2022} wherein the potential well associated with short-range electron transfer can be circumvented through long-range tunneling events. Spin relaxation was not measured in these scenarios, but we expect that the analogous short-range charge-transfer state, which is deleterious to free carrier generation, may also enhance the electron-hole exchange interactions that serve to hasten spin-valley relaxation in our bilayers.

\begin{figure*}
 \centering
 \includegraphics[height=9cm]{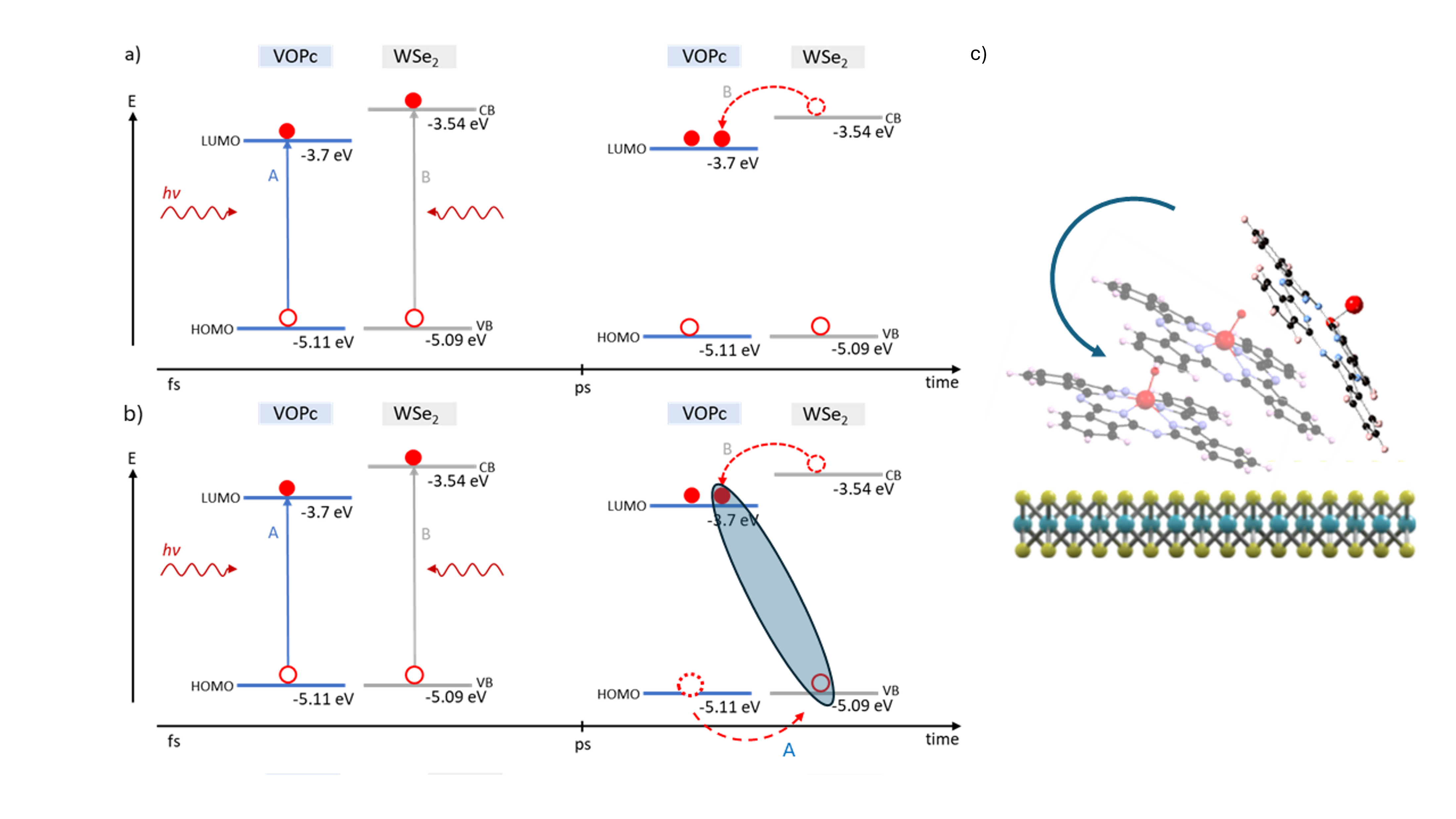}
 \caption{Schematic of the excited state processes in the VOPc / WSe$_2$ bilayers. a) Shows the dynamics observed in the 10 nm VOPc bilayers where the VOPc is oriented at 70$^\circ$ to the WSe$_2$. Photoexcitation of VOPc (pathway A) only leads to native VOPc kinetics, whereas excitation of WSe$_2$ (pathway B) incites electron transfer to VOPc. b) Shows the dynamics observed in the 5nm VOPc bilayers, associated with a flatter orientation of molecules with respect to WSe$_2$. The oval indicates the tightly-bound interfacial state that forms for both pathways A and B. c) Shows a schematic of the hypothesized change in molecular orientation with layer thickness, with the plane of the phthalocyanine becoming closer to parallel with the surface of WSe$_2$ as thickness is reduced.}
 \label{fgr:schematic}
\end{figure*}

Figure 5b summarizes the unique situation presented by the thinnest VOPc layer on WSe$_2$. Depicted as evolving to a more flat-lying geometry on the surface, we predict modified electronic structure that engenders a tightly-bound state, depicted with the blue oval, with distinct characteristics from that of the isolated components. Although perturbations are difficult to detect in the steady-state absorption, the new features in the TA bleach (Figure 4a) provides a verification of this interfacial state.  While molecule/TMDC states like this one are of fundamental interest,\cite{zhang_unraveling_2023} the concomitant destruction of spin-valley polarization relaxation is detrimental to potential spintronics\cite{yang_two-dimensional_2022, ahn_2d_2020} or QIS applications. 

 Ostensibly simpler than the bilayers, the spin-valley relaxation behavior of the bare WSe$_2$ monolayer remains to be fully explained. The fastest decay, associated with intervalley scattering in WSe$_2$, is mostly unchanged throughout the sample series, as one might expect.  The anomalous result is the behavior of the intermediate decay, which has not been definitively assigned a mechanism, but has been conjectured to involve trapped holes.\cite{Wagner_2021, liu2024} Our results concur with this notion, as the decay component is entirely absent from the 5 nm bilayer and then is relatively unchanged across the remainder of the thickness series.  If tight coupling of the 5 nm layer alters the effective WSe$_2$ surface electronic structure, isolated holes may be disfavored kinetically or energetically.  As thickness increases and the two layers decouple, this trapped hole pathway returns.

\section{Conclusions}
The work presented here provides evidence of distinct interaction regimes depending on the thickness of a molecular electron acceptor layer in a TMDC/molecular bilayer. Counter-intuitive quenching of the spin-valley polarization is observed for a 5 nm thick layer, whereas slightly thicker films of 7~nm and 10~nm VOPc show a prolongation of the spin-valley polarization onto the ns scale.  There is mounting evidence that thicker and more organized molecular layers produce the slowest spin-valley polarization relaxation concomitant with charges that are at least somewhat mobile. Studying even thicker molecular layers could prove useful for further separating charges and elongates the spin-valley polarization, but it creates practical issues in the WSe$_2$/VOPc system due to spectral overlap. Future work aimed at understanding the detailed mechanism of the charge transfer process and the nature of the interface state may reveal further design principles for enhancing the useful properties of these heterobilayers.

\begin{acknowledgement}
This work was authored by the National Renewable Energy Laboratory (NREL), operated by Alliance for Sustainable Energy, LLC, for the U.S. Department of Energy (DOE) under Contract No. DE-AC36-08GO28308. Bilayer sample characterization and analysis of spectroscopic data were supported by the Laboratory Directed Research and Development (LDRD) Director's Fellowship Program at NREL. Transient spectroscopy was supported by the Solar Photochemistry Program of the Department of Energy, Office of Basic Energy Sciences, Division of Chemical Sciences, Biosciences, and Geosciences. The views expressed in the article do not necessarily represent the views of the DOE or the U.S. Government. The U.S. Government retains and the publisher, by accepting the article for publication, acknowledges that the U.S. Government retains a nonexclusive, paid-up, irrevocable, worldwide license to publish or reproduce the published form of this work, or allow others to do so, for U.S. Government purposes.

\end{acknowledgement}

\begin{suppinfo}

Experimental procedures, sample characterization, and additional transient spectroscopic data.

\end{suppinfo}

\bibliography{rsc}

\end{document}